\documentstyle[aps]{revtex}
\input{epsf}
\twocolumn
\begin{document}

\title{ Polarization dynamics of femtosecond pulses propagating in air }
\author{ M.~Kolesik   }
\address{ Arizona Center for Mathematical Sciences,
        Department of Mathematics, University of Arizona,
        Tucson, AZ 85721. \\
        and \\
        Institute of Physics, Slovak Academy of Sciences,
        Bratislava, Slovakia
        }

\author{ J.V.~Moloney and E.M.~Wright }
\address{ Arizona Center for Mathematical Sciences,
        Department of Mathematics, University of Arizona,
        Tucson, AZ 85721. \\
        and \\
        Optical Sciences Center, University of Arizona,
        Tucson, AZ 85721.
        }
\date{\today}
\maketitle
\begin{abstract}
Polarization dynamics of femtosecond light pulses propagating in
air is studied by computer simulation. A rich variety of dynamics
is found that depends on the initial polarization state and power
of the pulse. Effects of polarization on the plasma and
supercontinuum generation are also discussed.
\end{abstract}

\section{Introduction}
There has been a great interest in long distance femtosecond pulse
propagation in air in recent years motivated in part by potential
applications in remote sensing and laser-induced lightning. The
first experimental observations\cite{BraKorLiu95,NibCurGri96} of
highly localized, high-intensity filaments propagating over
distances that exceed their corresponding Rayleigh lengths by
orders of magnitude motivated the efforts to understand the
phenomenon
\cite{BroChiIlk97,KosKanBro97,LanGriRip98,WosWedWil98,Fontaine99,Talebpour99}.
Due to the violent formation process, and the extreme time and
spatial scales of the filaments, the computer simulations and
analytic approaches turned out to be vital tools to grasp the
underlying physics
\cite{MleWriMol98,MleWriMol98b,MleWriMol99,MleKolWriMol99,Chiron99,Couairon00,Berge00,Akozbek00}.
Several models have been proposed.
The first explanation suggested a stationary waveguide
formed by competing effects of nonlinear self-focusing and linear
defocusing by underdense plasma generated in the most intense part of the
pulse\cite{BraKorLiu95}.
An alternative model employed a notion of the ``moving focus'' to
explain how the geometrical focus of a beam is transformed into 
a long filament\cite{BroChiIlk97}.
More recently, a dynamic spatial replenishment model emerged from
numerical investigation by Mlejnek et al.\cite{MleWriMol98,MleWriMol98b}.
The essential feature of the femtosecond propagation of infrared
(IR) pulses is its dynamics that makes it possible that the
localized filaments propagate over long distances and do not
suffer significant energy loses. The basic mechanism
involves a dynamic balance between the nonlinear self-focusing and
defocusing by free electrons generated via multi-photon
absorption by the high-intensity filaments. When the self-focusing
leading edge of the pulse starts to generate plasma, it thus
creates a defocusing ``lens'' for the trailing portion of the
pulse. This has two effects; First, it limits the losses due to
absorption in the plasma and, second, it prevents the major
portion of the pulse from experiencing self-focusing collapse.
After the plasma generating leading portion of the pulse exhausts
its energy, the strength of the defocusing lens decreases, and
the self-focusing starts over again, this time in the ``next
temporal slice'' of the pulse. The whole process can repeat
several times, depending on the total energy of the pulse. The
qualitative features of the dynamical spatial replenishment remain
valid also in transversely wide pulses, that break-up into
multiple filaments \cite{MleKolWriMol99}.

Up to the present, most of the work on the experimental side and
all numerical studies have been concerned with the case of
linearly polarized input pulses. Recently, Petit et al.
\cite{Petit00} studied the effects of the polarization on the
propagation of femtosecond IR pulses. They have measured
luminescence from the plasma generated in the filaments to show
that the polarization of the pulse plays an important role in the
plasma generation. Due to the highly dynamic nature of the
filament formation and propagation, it is natural to expect a rich
polarization dynamics in femtosecond pulses. In this initial
study, we restrict ourselves to femtosecond pulses with modest
peak powers that retain their initial radial transverse symmetry
and are just sufficient to produce several refocusing events
within a single pulse. The initial polarization of the pulse is
varied and the polarization state is recorded along the
propagation path. Our results indicate a tight correlation between
the evolution of the pulse waveform along the propagation distance
and certain global polarization parameters. Thus, the measurements
of the polarization state could provide yet another tool to
extract information on dynamics of pulses as well as an
opportunity to correlate experiment and theory.

The remainder of the paper is organized as follows. Our basic
model is described in Sec. II, and details of our simulated
experiment are given in Sec. III. Plasma production, polarization
dynamics, and the associated supercontinuum generation are then
discussed in Secs. IV-VI. Finally, our summary and conclusions are
given in Sec. VII.

\section{Model Equations}
Since the typical transverse dimension of the self-focusing
collapsing filaments are a few tens to hundred microns in the IR
wavelength region, it is a good approximation to consider the
optical field as transverse. We describe the complex optical
amplitude in terms of two circularly polarized components, ${\cal
E}^\pm$. The choice of the circularly polarized basis is motivated
by the fact that in this basis the nonlinear birefringence is
``diagonal'' and preserves the power in each component, which in
turn makes its implementation easier. Our model is a
straightforward extension of the scalar version we used in our
previous studies \cite{MleWriMol98,MleKolWriMol99}. It takes into
account the effects of diffraction, normal group velocity
dispersion (GVD), multi-photon ionization (MPI) and avalanche
plasma generation, defocusing of light by the generated free
electrons, and instantaneous and delayed cubic nonlinearity:
\begin{eqnarray}
\label{eqn:field}
 \frac{\partial {\cal E}^\pm}{\partial z}
=&&
 \frac{i}{2k}\nabla _{\perp}^{2}{\cal E}^\pm
-\frac{ik^{\prime \prime }}{2}
 \frac{\partial^{2}{\cal E}^\pm}{\partial t^{2}}  \\
-&&\frac{\sigma }{2}(1+i\omega \tau )\rho {\cal E}^\pm
-\frac{\beta ^{(K)}}{2}|{\cal E}|^{2K-2}{\cal E}^\pm \nonumber \\
+&& i\,\frac{2 \omega}{3 c} (1-f) n_{2}(|{\cal E}^\pm|^2 + 2 |{\cal E}^\mp|^2)  {\cal E}^\pm \nonumber \\
+i\,\frac{2 \omega}{3 c} f n_{2}
 &&\left[
     \int_{-\infty }^{\infty }dt^{\prime }R(t-t^{\prime })
        (|{\cal E}(t^{\prime })^\pm|^2 + 2 |{\cal E}(t^{\prime })^\mp|^2)
 \right] {\cal E}^\pm \ \  .\nonumber
\end{eqnarray}
Here $\omega$ is the optical frequency, $|{\cal E}|^2=|{\cal
E}^+|^2+|{\cal E}^-|^2$ the combined intensity of the two
circularly polarized components, $k=\omega/c$,
$k''=\partial^2k/\partial\omega^2$, $\rho$ is the electron
density, $\sigma$ is the cross-section for inverse bremsstrahlung,
$\tau$ is the electron collision time, $\beta^{(K)}$ is the
$K$-photon absorption coefficient, and the nonlinear change in
refractive-index for a linearly polarized continuous wave (cw)
field is $n_2|{\cal E}|^2$. The corresponding critical power for
self-focusing collapse for linearly polarized fields is
$P_{cr}=\lambda_0^2/2\pi n_2$, or $P_{cr}=$1.7~GW for our
parameters. In contrast, the critical power for self-focusing for
circularly polarized fields $P_{cr}^\pm=1.5 P_{cr}$ is 1.5 times
that for linearly polarized fields \cite{Marburger}. The
normalized response function (characterized by the resonance
frequency $\Omega$ and the decay $\Gamma$)
\begin{eqnarray}
R(t)&=&\theta(t)\;\Omega^2 \; e^{-\Gamma t/2}\;
                    \frac{\sin(\Lambda t)}{\Lambda}\;,\qquad
\Lambda=\sqrt{\Omega^2-\Gamma^2/4}\;,
\end{eqnarray}
accounts for delayed nonlinear effects, and $f$ is the fraction of
the cw nonlinear optical response which has its origin in the
delayed component, and we denoted the Heaviside step function by
$\theta(t)$. In the present model, we chose the relative weight of
the ``self'' and ``cross'' non-linear birefringent terms the same
as for the instantaneous Kerr effect and nonlinear Raman effect,
namely, the cross effect has a weight twice that of the self
effect as is appropriate for an isotropic medium \cite{Marburger}.

The optical field amplitude equations are completed by a simple
equation describing the evolution of the plasma density:
\begin{eqnarray}
\label{eqn:rho}
 \frac{\partial \rho }{\partial t}
&=&
 \frac{\sigma }{E_{g}}\rho |{\cal E}|^{2}
+\frac{\beta ^{(K)}|{\cal E}|^{2K}}{K\hbar \omega } -a\rho^{2}
\ ,
\end{eqnarray}
where the first and second terms represent the avalanche and
multi-photon effects, while the last one models the plasma
recombination. Note, that at the time-scales relevant for the
present study, the only practically important contribution comes
from the the multi-photon term. For the pulse powers we use, it is
sufficient to include oxygen alone as a source for MPI, since its
corresponding multiphoton order is lower than that for nitrogen.
We use the Keldysh theory formula to calculate the MPI rate
\cite{Keldysh}.

Explanations of symbols that appear in our model equations are
listed in Table 1 together with the values used in our simulation.

\section{Numerical simulation}
To study the role of the initial polarization state of the pulse
on its subsequent propagation, we performed a series of
simulations. In all runs, the initial pulse was chosen to be a
Gaussian wave-packet (in time and space) characterized through its
central wavelength $\lambda = 775$ nm, pulse waist $w = 0.7$ mm
and temporal duration $\tau_{\rm FWHM} = 200$ fs. We fixed the
initial peak intensity as $1.0\times 10^{16}$ W m$^{-2}$ which is a
relatively modest value: for a linearly polarized input this
corresponds to an input peak power of $P=7.7$ GW$=4.5 P_{cr}$, whereas for
a pure circular polarization $P=7.7$ GW$=3.0 P_{cr}^\pm$. At these
powers, there are typically two to three refocusing events in the
pulse propagation, and the numerics can be reliably controlled. At
higher powers under a perfect axial symmetry of the pulse, it may
be necessary to extend the model beyond the 
nonlinear Schr\" odinger equation
(NLSE), and to include
correction terms that allow to handle pulses with a very broad
spectrum. To check our numerical procedures and model
implementation, we performed most of the runs in two or three
different resolutions. The data we present were obtained with the
time-domain resolution of 0.24 fs. This resolution is sufficient
to capture spectra several hundredth of nanometers wide. We
provide more details of our testing procedures concerning the
spectral resolution and algorithm validity in Sec. VI devoted to
supercontinuum generation.

Below, we present our results for a series of runs that differ in
the initial polarization state of the pulse. We change the
polarization from the linear, through elliptic to circular, to see
how it affects the dynamics of the filaments. In previous work on
nonlinear propagation in fibers, the Stokes parameter formalism
has been employed to classify the polarization dynamics for plane
wave fields \cite{Daino86}. Here we employ space and time averaged Stokes
parameters $(s_0,s_1,s_2,s_3)$ as a description of the
polarization dynamics of the propagating pulses. In particular,
our Stokes parameters are calculated numerically according to the
prescription
\begin{eqnarray}
s_0 &=& ( {\cal F}_{\pi/2}  +  {\cal F}_{0}     )     \nonumber \\
s_1 &=& ( {\cal F}_{\pi/4}  -  {\cal F}_{-\pi/4})/s_0 \nonumber \\
s_2 &=& ( {\cal F}_{\pi/2}  -  {\cal F}_{0}     )/s_0 \nonumber \\
s_3 &=& ( {\cal F}_{\rm CR} -  {\cal F}_{\rm CL})/s_0
\end{eqnarray}
where ${\cal F}_\beta$ is the total energy detected after passing
the pulse through a polarizer of state $\beta$
\begin{equation}
{\cal F_\beta} = 2 \pi \int_{-\infty}^{+\infty} \int_0^R |{\cal
E}_\beta|^2(r,t) r dr dt
\end{equation}
with ${\cal E}_\beta$ the field resolved along the polarizer
direction, and $R$ the radius of a detection aperture chosen to
select the the most intense region of the pulse around the
filament. For a linear polarizer $\beta$ is the angle of the
polarizer, whereas for a circular polarizer $\beta=CR,CL$
corresponding to right and left circular polarization settings.
Here we chose $R=0.1$ mm, and perform the measurement of the
Stokes parameters ``in the near field''. We note that $s_0$ is the
total energy of the pulse detected over the aperture, and the
remaining Stokes parameters are calculated as differences between
the detected energy with different polarizer settings normalized
to the total energy. Thus our prescription has a direct
experimental interpretation and should therefore be of utility.
The above prescription for determining the Stokes parameters also
reduces to the usual definitions in the limit of long pulses of
broad transverse extent.

In the numerical simulations to be presented we fix the initial
Stokes parameter $s_2$ equal to zero, and vary $s_{1}$ and $s_3$
between zero and one to vary the initial polarization from linear
through elliptic to pure circular polarization. Besides the
polarization state, we also recorded the data pertaining to the
plasma generation inside the pulse, and generation of the
supercontinuum light. We start our discussion with plasma
generation.
\section{Plasma production}

\begin{figure}[t]
\centerline { \epsfxsize=3.5 in     \epsffile {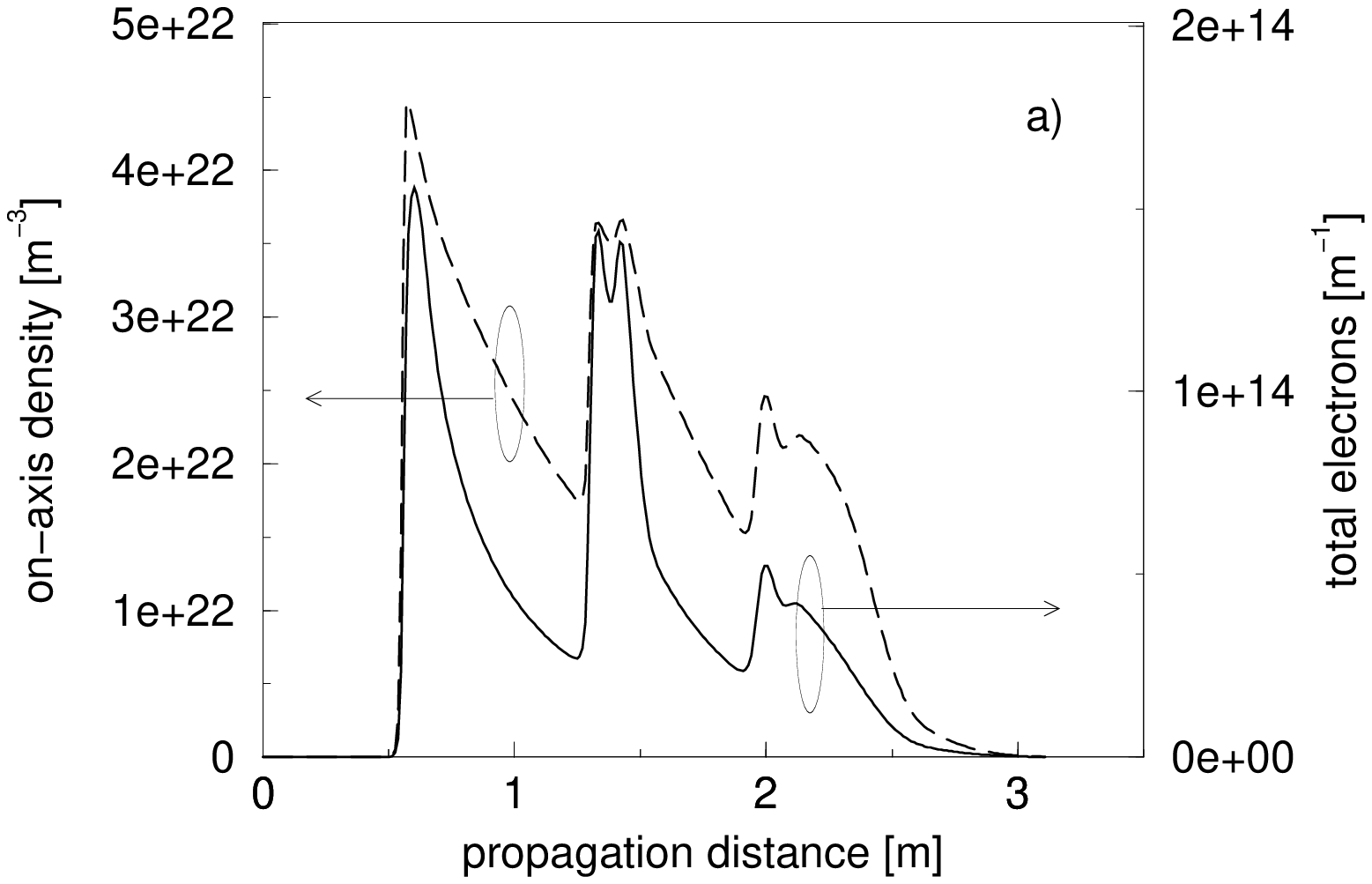} }
\centerline { \epsfxsize=3.5 in     \epsffile {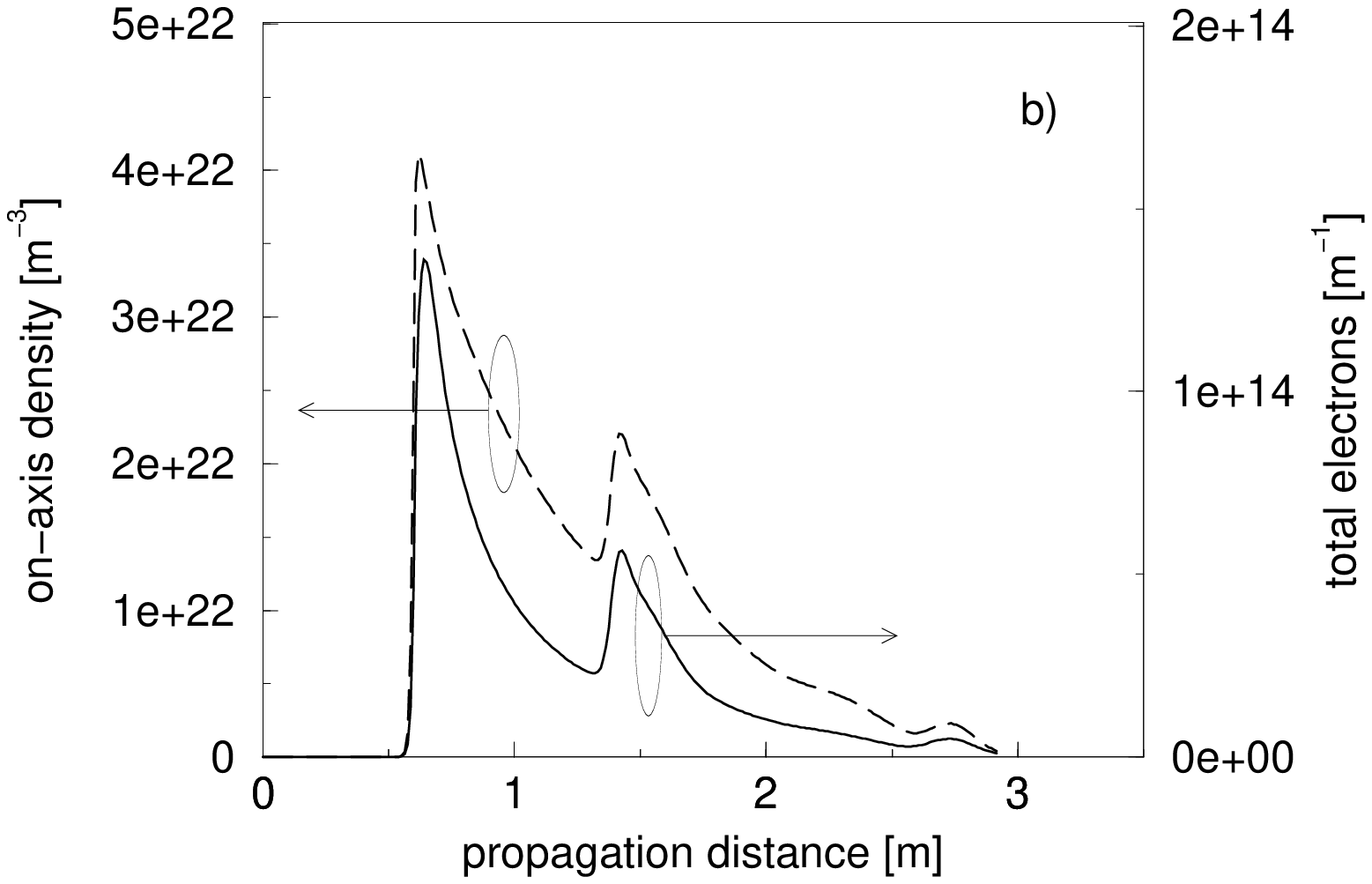} }
\centerline { \epsfxsize=3.5 in     \epsffile {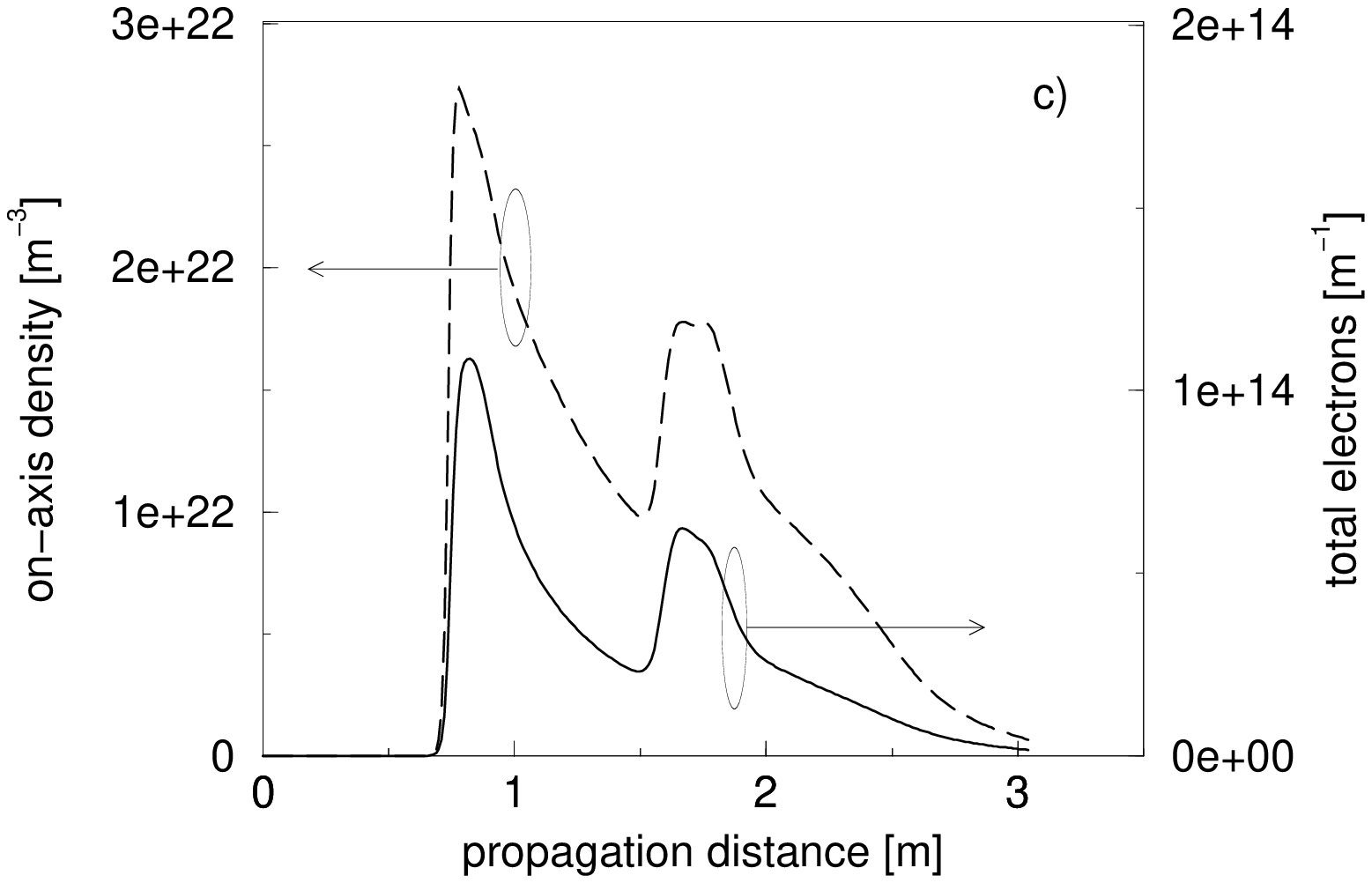} }
\caption{ Plasma generation for almost linear (a),
elliptic (b) and a close-to-circular (c)
initial pulse polarization. The total peak power is kept the same
in all cases. Since the critical power for self-focusing is higher
for circular polarization, the circularly polarized pulse 
experiences weaker self-focusing which in turn results in less
overall plasma generation. \label{fig:plasma} }
\end{figure}

\begin{figure}[t]
\centerline { \epsfysize=2.5 in     \epsffile {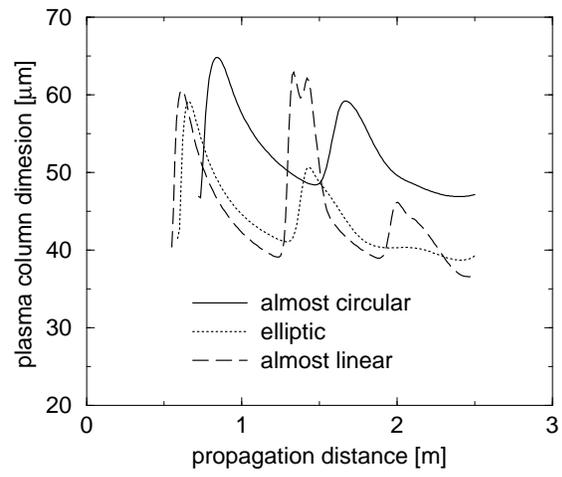} }
\caption{
Characteristic transverse dimension of the generated
plasma column for three different initial polarizations.
\label{fig:pladim}
}
\end{figure}

As the femtosecond pulse undergoes multiple self-focusings, the
amount of plasma generated by its high-intensity portions reflects
the spatio-temporal shape of the pulse. The total number of
generated electrons as well as the maximal plasma densities
exhibit peaks along the propagation distance. These peaks coincide
with the locations of self-focusing collapses within the pulse, each
peak being produced by a different temporal portion of the pulse.
Figure~\ref{fig:plasma} shows the plasma generation for three
different initial polarizations of the pulse. The trend that one
can see is quite in line with what is expected based on the
functional form of the nonlinear birefringence. Namely, as we
change the initial polarization from linear through elliptic to
circular, the amount of the generated plasma decreases. Also, the
onset of filament formation is delayed for the circularly
polarized pulse because the critical power for self-focusing is
higher for circularly polarized pulse as noted earlier. In other
words, keeping the input peak power the same for different
polarizations, we effectively decrease the self-focusing power of
circularly polarized pulses. This is also the reason why the
number of refocusing events can be higher in a close-to-linear or
linear polarization than in a circularly polarized pulse. While
the overall plasma production depends on the polarization state,
the typical dimensions of the filaments are not very sensitive to
it. That can be seen from the Figure~\ref{fig:plasma} which shows
the longitudinal extent of the plasma columns. The transverse
dimensions of the plasma channels can be estimated from the ratio
of the two curves shown in the figure as the square root of the
ratio between the total number of electrons and the maximal plasma
density. This characteristic dimension of the plasma channel is
shown in Fig.~\ref{fig:pladim} for three different polarizations.
Though there are small variation between different initial
polarizations, the thickness of the plasma channels is always
roughly 60 microns in the most dense parts. The plasma channel
generated by the circularly polarized pulse seems to be more
``homogeneous'', exhibiting less thickness variation along the
propagation distance.

\section{Polarization dynamics}
\begin{figure}[t]
\centerline { \epsfysize=2.5 in     \epsffile {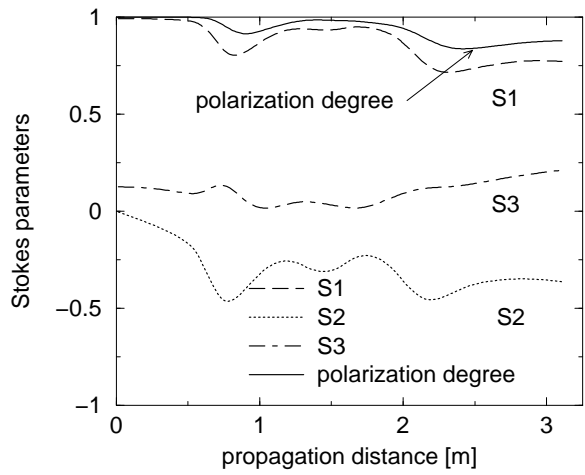} }
\caption{
Stokes polarization parameters of the on-axis part of the pulse
as functions of the
propagation distance for an almost linear initial
polarization. The initial deviation from
the perfect linear polarization increases, and the
polarization degree decreases.
\label{fig:pollin}
}
\end{figure}

\begin{figure}[t]
\centerline { \epsfysize=2.5 in     \epsffile {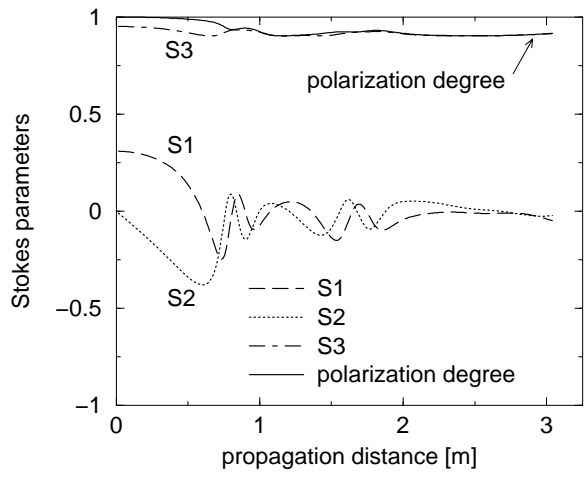} }
\caption{
An initially almost circularly polarized pulse (approximately)
preserves its polarization state. There is only a slight decrease of
the polarization degree.
\label{fig:polcir}
}
\end{figure}

\begin{figure}[t]
\centerline { \epsfysize=2.5 in     \epsffile {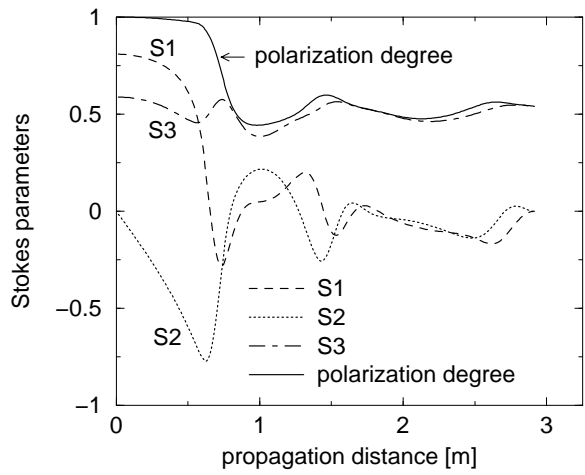} }
\caption{ Stokes polarization parameters as functions of the
propagation distance for an elliptic initial polarization of the
pulse. The central part of the filament evolves into a
predominantly circular polarization state after the second
self-focusing collapse event. Note that the rate of change of the
polarization state correlates with the loci of maximal plasma
production (see Fig.~\ref{fig:polrate}). \label{fig:poleli} }
\end{figure}

While the dynamics of the plasma generation and its dependence on
the polarization described in the previous Section is
straightforwardly linked to the structure of the equations
governing the optical field evolution, the polarization dynamics
seems to be more difficult to interpret.
Figures~\ref{fig:pollin},\ref{fig:polcir},\ref{fig:poleli} show
the Stokes parameters and the polarization degree as functions of
the propagation distance for the three different initial
polarizations we discussed in the previous Section. An interesting
feature is the difference between the ``stability'' of initial
linear and circular polarization. Figure~\ref{fig:pollin} shows
that a small perturbation to the linear polarization in the
initial pulse leads to an increasing deviation of the polarization
state from the initial one. In this sense, the linear polarization
appears to be unstable, as the polarization measured after the
filament formation can significantly differ from the initial one.
Naturally, the rate of divergence for two close but not identical
initial conditions decreases with the decreasing input power. On
the other hand, in the case of almost circular polarization shown
in Fig.~\ref{fig:polcir}, the pulse polarization state doesn't
change that dramatically. Though there is a small decrease of the
polarization degree, one can say that final polarization stays
close to the initial one even after two refocusings of the pulse.
Thus, the circular polarization seems to be more stable against
small perturbations than the linear polarization.
Figure~\ref{fig:poleli} shows an interesting case of an initially
elliptic polarization. Note that the Stokes parameter $s_3$, which
measures the degree of circular polarization, only exhibits small
variations, while the other two parameters decrease significantly
after their initial increase in the first collapse. That means
that the light focused in the second collapse is predominantly
circularly polarized. This observation is confirmed by examining
the spatio-temporal polarization pattern within the pulse. This is
an observation that may not be expected based on the previous
results concerning cw self-focusing of polarized pulses
\cite{Shen,Marburger}. Namely, in a situation close to a
continuous wave regime, one can argue that the weaker circular
component experiences a stronger focusing ``lens'' because of the
factor two in the birefringence cross-term, and that eventually
leads to equal intensities of both circular components and,
therefore, linear polarization of the central filament. However,
the important point here is that the femtosecond light filaments
under consideration are extremely dynamic objects. The resulting
polarization distributions strongly depend on the spatial and
temporal location within the pulse, and any interpretation based
on steady-state-like considerations becomes invalid. Namely, there
is a delay between when the light encounters the focusing ``lens''
and when it actually reaches the focus. This delay interferes with
the temporal profile of the pulse, which typically exhibits
multiple peaks that may be just a few femtosecond long. As a
consequence, the above simple argument is not sufficient to capture
all essential features of the phenomenon. The tendency of the
predominantly circular polarization of the most intense portion of
a filament was also observed in our simulations that were not
restricted to axial symmetry \cite{KolSPIE}. A wide beam with a random
perturbation breaks up into multiple filaments that exhibit
polarization properties similar to those we discuss here. However,
one has to keep in mind, that in both cases, axisymmetric as well
as fully spatially resolved, our simulation modeled pulses with
relatively small energy fluence when compared to some current
experiments. Therefore, it would be extremely interesting to see
what happens to the polarization of the central filament in a
pulse that has enough energy for many self-focusing events
and also has the transverse profile clean enough to preserve
its axial symmetry.

\begin{figure}[t]
\centerline { \epsfysize=2.5 in     \epsffile {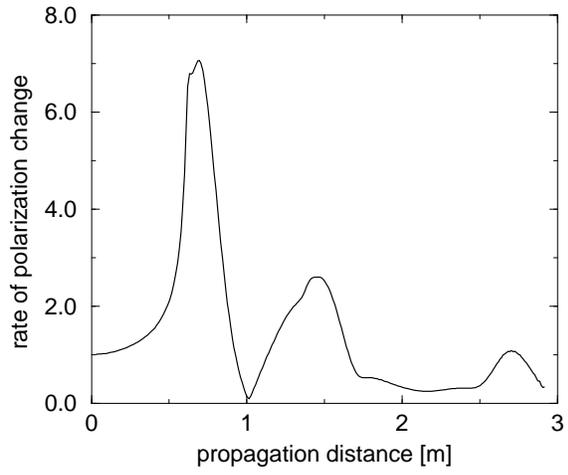}
} \caption{ \label{fig:polrate} Root-mean-square of the rate of
change of the Stokes polarization vector, $ds/dz$ (in units of
m$^{-1}$) as a function of the propagation distance for elliptic
initial polarization. The rate
maxima are correlated with the locations of strongest plasma
generation and focusing. }
\end{figure}

We conclude this Section with yet another presentation of the
polarization dynamics data we have shown above. Namely, we want to
show that the polarization changes closely reflect the
self-focusing events within the pulse and, consequently, the
locations where most of the plasma is generated.
Figure~\ref{fig:polrate} shows the root-mean-square rate of the
change of the Stokes vector along the propagation distance
\begin{equation}
\frac{ds}{dz} = \sqrt{\left (\frac{ds_1}{dz}\right )^2+\left
(\frac{ds_2}{dz}\right )^2 + \left (\frac{ds_3}{dz}\right )^2 }
\end{equation}
for the case of elliptic initial polarization of the pulse. The
curve shown corresponds to the data depicted in
Fig.~\ref{fig:poleli} and in
Fig.~\ref{fig:plasma}b). Note, that the maxima of the rate of the
polarization change closely follow those in the plasma production
curve. We thus see that the multiple self-focusing events in the
single pulse leave their signature on the polarization. This could
provide another way, besides the indirect plasma density
observations\cite{Talebpour99,Schill99,Chien00,Tzortzakis99}, 
to visualize the dynamics of the
spatial replenishment.

\section{Supercontinuum generation}

After contrasting the behavior of pulses polarized close to linear
and circular from the points of view of plasma generation and of
their polarization dynamics, we want to discuss the effects of
polarization on the supercontinuum generation. However, before
presenting our results, we feel a note concerning some technical
questions is in order. The explosive spectral broadening in the
supercontinuum generation in femtosecond pulses is a rather
subtle phenomenon from the point of view of numerical simulation.
Clearly, one needs a sufficient resolution in the time (spectral)
domain to capture the broad spectrum, but the resolution may not
be the only issue here. One has to check how broad is the spectral
region within which the model and its numerical implementation
describes the wave propagation correctly. It is expected, that at
extreme powers, correction terms beyond the basic NLSE (see e.g.
Ref.~\cite{Brabec97}) need to be included in the field equation.
To ensure that we work in the regime where the correction terms may
be neglected, and to assess the spectral band over which our
numerics works well, we performed some comparative simulations.
The choice of the reference frequency (wavelength) around which
the NLSE is built is in principle arbitrary, though it is
obviously most appropriate to choose it close to the central
frequency of the modeled pulse. This means, that simulations that
only differ in the choice of the reference frequency should give
the same results. We have compared simulation with the reference
frequency shift of 150 nanometers off the central wavelength of
the pulse, and obtained a very close match of the spectra in the
region from 500 to 1200 nm and over four decades in spectral
intensity. Thus, in this interval we can trust the spectra
extracted from our simulations. We would like to point out that this
is in fact a rather strong test for the overall numerical
implementation of the solver. But most important, the close
agreement between the spectra 
shows that the correction terms beyond NLSE do not play an important role in our
regime of modest input powers; This is because one can interpret
them as corrections that partially restore the invariance of the
original wave equation with respect to the choice of the
(physically meaningless) reference frequency -- that is exactly
what our test shows is not needed for the conditions in our
simulations.

\begin{figure}[t]
\bigskip
\centerline { \epsfysize=2.5 in     \epsffile {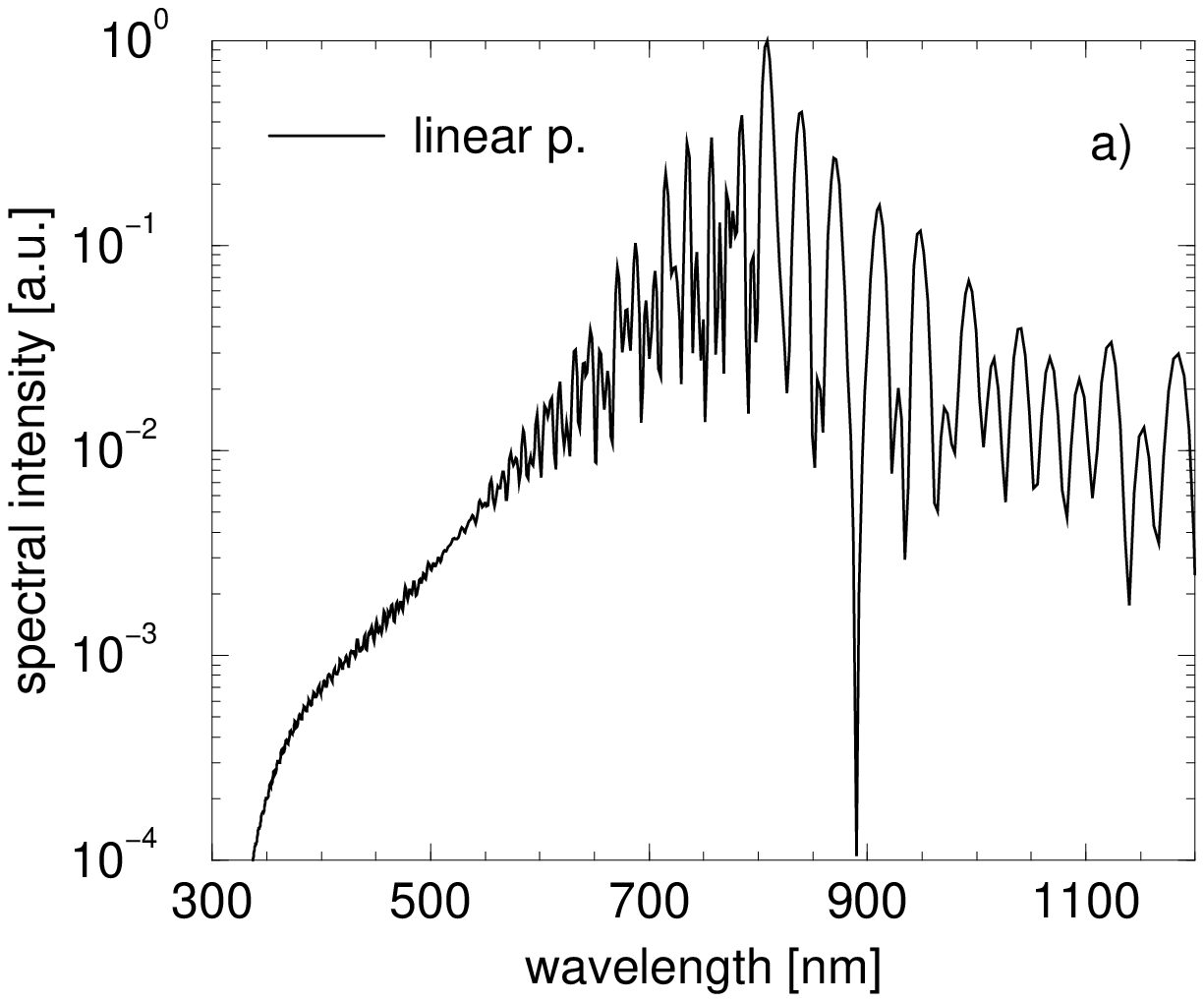} }
\centerline { \epsfysize=2.5 in     \epsffile {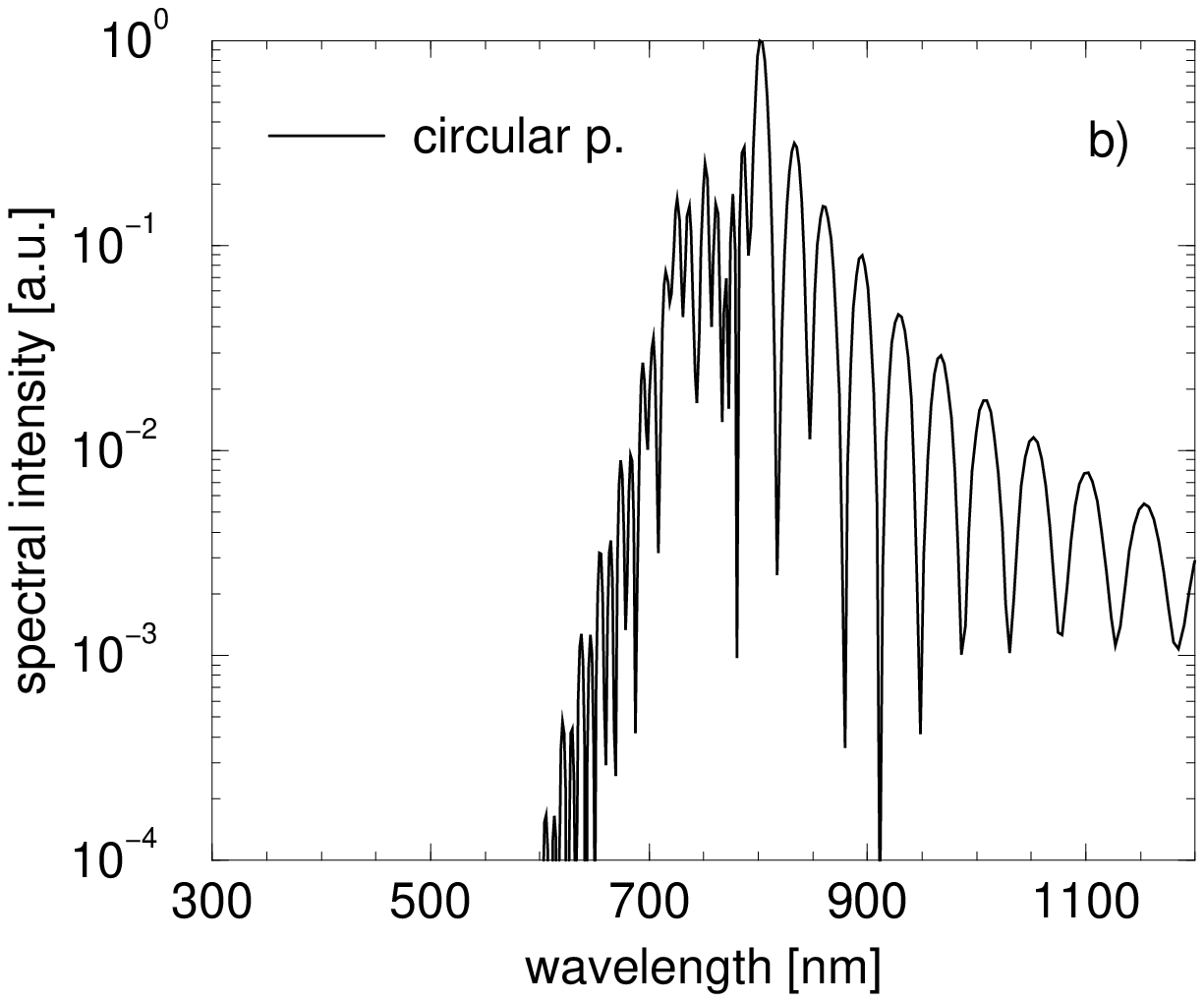} }
\caption{
Spectral broadening of two femtosecond pulses with different
initial polarizations. The linearly polarized pulse produces
significantly more supercontinuum light than an equally intense
circularly polarized pulse.
\label{fig:spccmp}
}
\end{figure}

Figure~\ref{fig:spccmp} shows comparison of the spectral
broadening of two pulses that differ only in their initial
polarization. The way we extracted the spectra from the pulse
waveforms corresponds to measurement in a near field with the same
aperture we used for polarization characterization. The figure
shows spectra ``measured'' after the last self-focusing collapse,
after which the pulse will eventually diffract and there will not
be more supercontinuum generation. One can see that the pulse
which was initially polarized close to linear exhibits a much
stronger supercontinuum generation. However, note that we compare
pulses with the same peak power, and what we see here is an effect
similar to the plasma generation. Supercontinuum generation
strongly depends on the available power and the natural measure of
that power is in units of critical power for self-focusing. From
that point of view the circularly polarized pulse is weaker and
that is the main reason that it exhibits less spectral broadening.
However, this situation represents a reasonable experimental setup
in which only the polarization is changed. 

To get a feeling
regarding the role of the group velocity dispersion in the
supercontinuum generation, we performed most of the simulation
runs also with a higher group velocity dispersion parameter. It
turns out that increasing GVD by an order of magnitude leads to a
strong suppression of the continuum production. We speculate that
it may be one of the reasons that, at least in some experiments,
there is only little spectral broadening in the ultraviolet (UV)
femtosecond pulses\cite{Diels} because the GVD value of air is
significantly higher in the UV region.

The findings from our numerical simulations should be accessible
to experimental testing. However, extreme caution should be
exercised when trying to compare experimental and simulational
spectra. The spectra we present are taken from ``one shot''. They
exhibit modulations typical for supercontinuum generation in gases
\cite{NibCurGri96,Corkum}. While the characteristic ``frequency'' of the
modulation is rather reproducible, the exact spectral shape is
not. Thus, even small fluctuation in the parameters of the pulses
will result in a suppression of fine features in the multiple-shot
experimental spectra.

\section{Conclusions }

We have performed a computer simulation study of the
effects of the initial pulse polarization on its
propagation and filamentation dynamics.
In agreement with the experiment~\cite{Petit00}, we have found that the
filamentation onset is reached earlier for a linearly
polarized pulse than in a circularly polarized pulse
of the same peak power. However, in some cases, the
experiment indicates that the circularly polarized
pulses create higher plasma densities in comparison with
linearly polarized pulses. Our simulations suggest
the opposite, but one has to keep in mind that the
experimental measurements and our simulation pertain
to rather different conditions, including much higher
power and focusing in the experiment. In our simulations,
we also see more self-focusing events with linearly
polarized pulses than with circularly polarized ones.

We have observed a rich spatio-temporal polarization dynamics.
Naturally, the limited range of the full parameter space explored
in our simulations prevents us from drawing general conclusions,
but we believe some tendencies are already discernible. First, the
initially circular polarization seems to be stable in a sense that
a small perturbation of the polarization state doesn't grow
significantly. On the other hand, when a small polarization
perturbation is applied to the linearly polarized pulse, it grows
and the polarization degree of the central filament decreases
significantly. The growth rate of the deviation is expected to
increase for higher powers. An interesting case is the one of an
initially elliptic polarization. We have observed that the center
of the filament is almost purely circularly polarized after
subsequent self-focusing collapses within the pulse. Apparently,
we have here a rather different situation than in the
self-focusing in nanosecond pulses, which tend to create linearly
polarized filaments independently of initial polarization
state\cite{Shen,Marburger}.
The usual argument, that the the weaker circular component
experiences stronger self-focusing which finally leads to
balancing the power of the circularly polarized components
and thus to linear polarization, can't be applied
in the femtosecond pulses. The evolution of the polarization
state along the propagation distance is extremely dynamic
and hardly possible to describe in simple ``static'' terms.
The multiple refocusing within a pulse and the
defocusing effect of the generated plasma play a major role.
The important feature of the whole process is that apart
from the relatively small energy losses due to plasma generation
and radiation, the energy in each circular component remains
conserved. Thus, the main mechanism that results in
changing polarization pattern within the pulse is the spatio-temporal
energy redistribution within each circular component.
As a consequence, the polarization state of the whole pulse
is very complicated and, therefore, any projection onto
global quantities like Stokes parameters has to be interpreted
in relation to details of the measurement (aperture, near vs.
far field, collecting angle, \ldots ).

We have also looked at spectra ``measured'' after the last
self-focusing collapse for different initial polarizations.
In accordance with our observation about the plasma generation,
we see much stronger supercontinuum generation in linearly
polarized than in circularly polarized pulses.

Finally, we have seen that rate of change of the polarization
state in the center of the filament is closely correlated
with self-focusing and plasma generation. Thus, the polarization
offers, in principle, an alternative way to investigate
the dynamics of the spatial replenishment in femtosecond pulses.

In this work, we have concentrated on investigating ``global''
quantities to characterize the femtosecond pulse propagation that
should be experimentally accessible, at least in principle. Naturally,
the question is how much can be done practically. To measure the
Stokes parameters evolution along the propagation distance,
for example,
a very good reproducibility of the initial pulse would be required.
However, the ``final output'' polarization state and spectra,
which should be easier to measure, also carry a lot of signatures
about the inner dynamics of the femtosecond pulse propagation.

\section{Acknowledgments}
Work supported by AFOSR grant no. F4962-00-1-0312, AFOSR DURIP
grant no. F4962-00-1-0190, and in part by Boeing. M.K. was partly
supported by the GASR grant VEGA 2/7174/20.


\begin{thebibliography}{10}

\bibitem{BraKorLiu95}
A. Braun {\it et~al.}, Opt. Lett. {\bf 20},  73  (1995).

\bibitem{NibCurGri96}
E. Nibbering {\it et~al.}, Opt.Lett. {\bf 21},  62  (1996).

\bibitem{BroChiIlk97}
A. Brodeur {\it et~al.}, Opt. Lett. {\bf 22},  304  (1997).

\bibitem{KosKanBro97}
O. Kosareva {\it et~al.}, Opt.Lett. {\bf 22},  1332  (1997).

\bibitem{LanGriRip98}
H. Lange {\it et~al.}, Opt. Lett. {\bf 23},  120  (1998).

\bibitem{WosWedWil98}
L. {W\"oste} {\it et~al.}, AT-Fachverlag, Stuttgart, Laser und Optoelektronik
  {\bf 29},  51  (1997).

\bibitem{Fontaine99}
B. La~Fontaine {\it et~al.}, Phys. of Plasmas {\bf 6},  1615  (1999).

\bibitem{Talebpour99}
A. Talebpour, S. Petit, and S. Chin, Opt. Commun. {\bf 171},  285  (1999).

\bibitem{MleWriMol98}
M. Mlejnek, E. Wright, and J. Moloney, Opt. Lett. {\bf 23},  382  (1998).

\bibitem{MleWriMol98b}
M. Mlejnek, E. Wright, and J. Moloney, Phys. Rev. E {\bf 58},  4903  (1998).

\bibitem{MleWriMol99}
M. Mlejnek, E. Wright, and J. Moloney, IEEE J. Quant. Electron. {\bf 35},  1771
   (1999).

\bibitem{MleKolWriMol99}
M. Mlejnek, M. Kolesik, J. Moloney, and E. Wright, Phys. Rev. Lett. {\bf 83},
  2938  (1999).

\bibitem{Chiron99}
A. Chiron {\it et~al.}, Eur. Phys. J. D {\bf 6},  383  (1999).

\bibitem{Couairon00}
A. Couairon and L. Berg\'e, Phys. of Plasmas {\bf 7},  193  (2000).

\bibitem{Berge00}
L. Berg\'e and A. Couairon, Phys. of Plasmas {\bf 7},  210  (2000).

\bibitem{Akozbek00}
N. Ak\"ozbek, C. Bowden, A. Talebpour, and L. Chin, Phys. Rev. E {\bf 61},
  4540  (2000).

\bibitem{Petit00}
S. Petit, A. Talebpour, A. Proulx, and S. Chin, Opt. Commun. {\bf 175},  323
  (2000).

\bibitem{Marburger}
J. Marburger,  in {\em Prog. Quant. Elect.} (Pergamon Press, Oxford, 1977),
  Vol.~4, p.\ 35.

\bibitem{Keldysh}
L. Keldysh, Sov. Phys. JETP {\bf 20},  1307  (1965).

\bibitem{Daino86}
B. Daino, G. Gregori, and S. Wabnitz, Opt. Lett. {\bf 11},  42  (1986).

\bibitem{Shen}
Y. Shen,  in {\em Prog. Quant. Elect.} (Pergamon Press, Oxford, 1977), Vol.~4,
  p.\ 1.

\bibitem{KolSPIE}
M. Kolesik, M. Mlejnek, J. Moloney, and E. Wright,  in {\em Optical pulse and
  beam propagation II}, Vol.~3927 of {\em Proceedings of SPIE}, edited by Y.
  Band (SPIE, Bellingham, 2000), p.\ 81.

\bibitem{Schill99}
H. Schillinger and R. Sauerbrey, Appl. Phys.B {\bf 68},  753  (1999).

\bibitem{Chien00}
C. Chien {\it et~al.}, Opt. Lett. {\bf 25},  578  (2000).

\bibitem{Tzortzakis99}
S. Tzortzakis {\it et~al.}, Phys. Rev. E {\bf 60},  R3505  (1999).

\bibitem{Brabec97}
T. Brabec and F. Krausz, Phys. Rev. Lett. {\bf 78},  3282  (1997).

\bibitem{Diels}
J. Schwartz {\it et~al.}, Opt. Commun. {\bf 180},  383  (2000).

\bibitem{Corkum}
P. Corkum and C. Rolland,  in {\em The supercontinuum laser source}, edited by
  R. Alfano (Springer, New York, 1989), p.\ 318.

\end{thebibliography}

\newpage

\onecolumn
\narrowtext

\newpage

\newpage

\newpage

\newpage

\newpage

\newpage

\newpage

\begin{table}
\caption{Model parameters and numerical values used in our simulations.}
\label{tab:params}
\begin{tabular}{|c|c|l|}
Quant.& Value\&Unit &Note \\ \tableline
$k$& $k=2 \pi/\lambda_0$  & reference wavevector \\
$\lambda_0$& $775\times10^{-9}$ m & wavelength \\
$k''$ & $2.1\times10^{-29}$ s$^2$/m & group velocity dispersion \\
$n_2$& $5.6\times10^{-19}$ cm$^2$/W & nonlinear index \\
$f$& $0.5$   &   \\
$\Gamma$ & $26$ THz & $R(t) \sim \theta(t) \times$ \\
$\Lambda$ & $16$  THz & \hspace{0.1in}
                              $e^{-\Gamma t/2}\,\sin(\Lambda t)$ \\
$K $& 7  & MPI order \\
$E_g $& $\approx 11$ eV & ionization energy \\
$\beta_{K}$& $6.5\times10^{-104}$ m$^{11}$W$^{-6}$ & MPI rate \\
$\tau$& $3.5\times10^{-13}$ s & electron collision time \\
$\sigma$
& $5\times10^{-24}$ m$^2$ & cross-section for \\
&& inverse bremsstrahlung \\
$a$& $5\times10^{-13}$ m$^3$/s & recombination rate \\
\end{tabular}
\end{table}

\end{document}